\documentclass[pra,aps,onecolumn,nopacs,superscriptaddress,nofootinbib]{revtex4}
\usepackage[T1]{fontenc}
\usepackage[latin1]{inputenc}
\usepackage{lmodern}\usepackage{graphicx}
\usepackage{dcolumn}
\usepackage{bm}
\usepackage{amsmath}
\usepackage{amssymb}
\usepackage{pst-all}
\usepackage{psfrag}
\usepackage{epsfig}

\def\ii{{\rm i}}     
\def\Rb{{\bf R}}  \def\pb{{\bf p}}    
    \def\Eb{{\bf E}}  \def\rb{{\bf r}}

\def\EF{{E_{\rm F}}}  \def\vF{{v_{\rm F}}}

\begin{document}

\title{Molecular Sensing with Tunable Graphene Plasmons}

\author{Andrea~Marini}
\affiliation{ICFO-Institut de Ciencies Fotoniques, Mediterranean Technology Park, 08860 Castelldefels (Barcelona), Spain}
\email{andrea.marini@icfo.es}
\author{Iv\'an~Silveiro}
\affiliation{ICFO-Institut de Ciencies Fotoniques, Mediterranean Technology Park, 08860 Castelldefels (Barcelona), Spain}
\author{F.~Javier~Garc\'{\i}a~de~Abajo}
\affiliation{ICFO-Institut de Ciencies Fotoniques, Mediterranean Technology Park, 08860 Castelldefels (Barcelona), Spain}
\affiliation{ICREA-Instituci\'o Catalana de Recerca i Estudis Avan\c{c}ats, Barcelona, Spain}
\email{javier.garciadeabajo@icfo.es}

\begin{abstract}
We study the potential of graphene plasmons for spectrometer-free sensing based on surface-enhanced infrared absorption and Raman scattering. The large electrical tunability of these excitations enables an accurate identification of infrared molecular resonances by recording broadband absorption or inelastic scattering, replacing wavelength-resolved light collection by a signal integrated over photon energy as a function of the graphene doping level. The high quality factor of graphene plasmons plays a central role in the proposed detection techniques, which we show to be capable of providing label-free identification of the molecular vibration fingerprints. We find an enhancement of the absorption and inelastic scattering cross-sections by 3-4 orders of magnitude for molecules in close proximity to doped graphene nanodisks under currently feasible conditions. Our results pave the way for the development of novel cost-effective sensors capable of identifying spectral signatures of molecules without using spectrometers and laser sources.
\end{abstract}
\maketitle

\section{Introduction}

Structural vibrations in molecules produce infrared spectral features that can be regarded as specific barcodes, therefore allowing us to resolve their chemical identity. However, because the molecules are much smaller than the optical wavelength, their interaction with light is extremely weak. Fortunately, the tight confinement and large field enhancement produced by plasmons --the collective excitations of conduction electrons in metals-- offer a solution to increase this interaction. By exposing the target molecules to the plasmons of metallic nanostructures, they greatly improve their ability to absorb and inelastically scatter light. This is the underlying principle of the techniques known as surface-enhanced infrared absorption (SEIRA) and surface-enhanced Raman scattering (SERS) \cite{M1985,M05_2}. In a complementary direction, the substantial spectral shifts experienced by plasmons upon adsorption of molecular layers has been also used for sensing both in metallic colloids \cite{HYG99,WV07,AHL08} and in lithographically prepared structures \cite{AKG09,CA12_2}, although this approach needs to be combined with markers in order to resolve the molecular identity. In contrast, SERS and SEIRA enable full characterization of the roto-vibrational molecular structure (i.e., the fingerprint of the adsorbed molecules), with a sensitivity that goes down to the single molecule detection limit \cite{KWK97,NE97,paper125}. These techniques have already enabled a number of viable applications (e.g., pregnancy tests based on metal colloids \cite{Leuvering} and cancer screening \cite{LFH14}), while their great potential has generated expectations for revolutionary applications in plasmonic sensing. Nevertheless, there are some aspects on which further improvement should help: metal plasmons lack post-fabrication external tunability, with their frequencies essentially determined by composition, geometry, and environment; additionally, the spectral width of each individual plasmon is limited to a narrow range of infrared frequencies, thus enhancing only a few of the molecular resonances. In this respect, elaborate hole array \cite{LDD13} and nanoantenna \cite{ASR13} designs have been used with some success to cover a wider spectral region. 

Doped graphene has recently emerged as an attractive alternative to noble metals, as it shows electrically tunable surface plasmons at infrared (IR) and THz frequencies \cite{WSS06,HD07,JGH11,FAB11,SKK11,paper196,FRA12,YLC12,YLL12,paper212,BJS13,YLZ13,paper230}. The electronic band structure of this material is characterized by a combination of linear dispersion relation (i.e., uniform Fermi velocity $\vF\approx10^{6}$\,m/s) and vanishing density of states at the Fermi level in its neutral state \cite{W1947,CGP09}. Besides, graphene exhibits metallic optical response when its Fermi energy $\EF$ is moved away from the neutrality point, leading to the existence of plasmons. Broad evidence of graphene plasmons and their electrical tunability has been experimentally gathered \cite{JGH11,FAB11,SKK11,paper196,FRA12,YLC12,YLL12,paper212,BJS13,YLZ13,paper230}, while the main features of these excitations are theoretically well understood in both extended \cite{WSS06,HD07} and structured \cite{BF07,JBS09,VE11,paper176,NGG11,paper181,paper182,paper194,paper212,BPV12,FP12,NGG12,WK13,paper235,paper214,SHG14,paper237} atomically thin carbon films. These studies have revealed some unique properties of graphene plasmons, including their long lifetimes, large spatial confinement and field enhancement, and extraordinary tunability via electrostatic gating. Besides, graphene is a very nonlinear material \cite{M07_2,HHM10,M11} whose plasmons have been predicted to produce extraordinary nonlinear effects \cite {DCM13,SK14,SSS14,SHG14,paper247}. Moreover, the interest in graphene has spurred the design of sensors in which propagating plasmons produce enhancement of the molecular vibrational features \cite{FGY14}, the graphene acts as a cleaner surface \cite{XLX12}, or the plasmons are shifted by the adsorption of molecules \cite{WCK10,LYF14}.

Here, we explore the potential of localized graphene plasmons for sensing via SEIRA and SERS. Based upon realistic numerical simulations of currently attainable graphene structures, we show that the recorded signal integrated over a broadband spectral range is sufficient to provide chemical identification when it is examined as a function of $\EF$. This constitutes a solid basis to support graphene as an ideal platform for spectrometer-free sensing, which is enabled by the extraordinary electrical tunability of this material.

\section{Results and discussion}

\begin{figure*}
\begin{center}
\includegraphics[width=160mm,angle=0,clip]{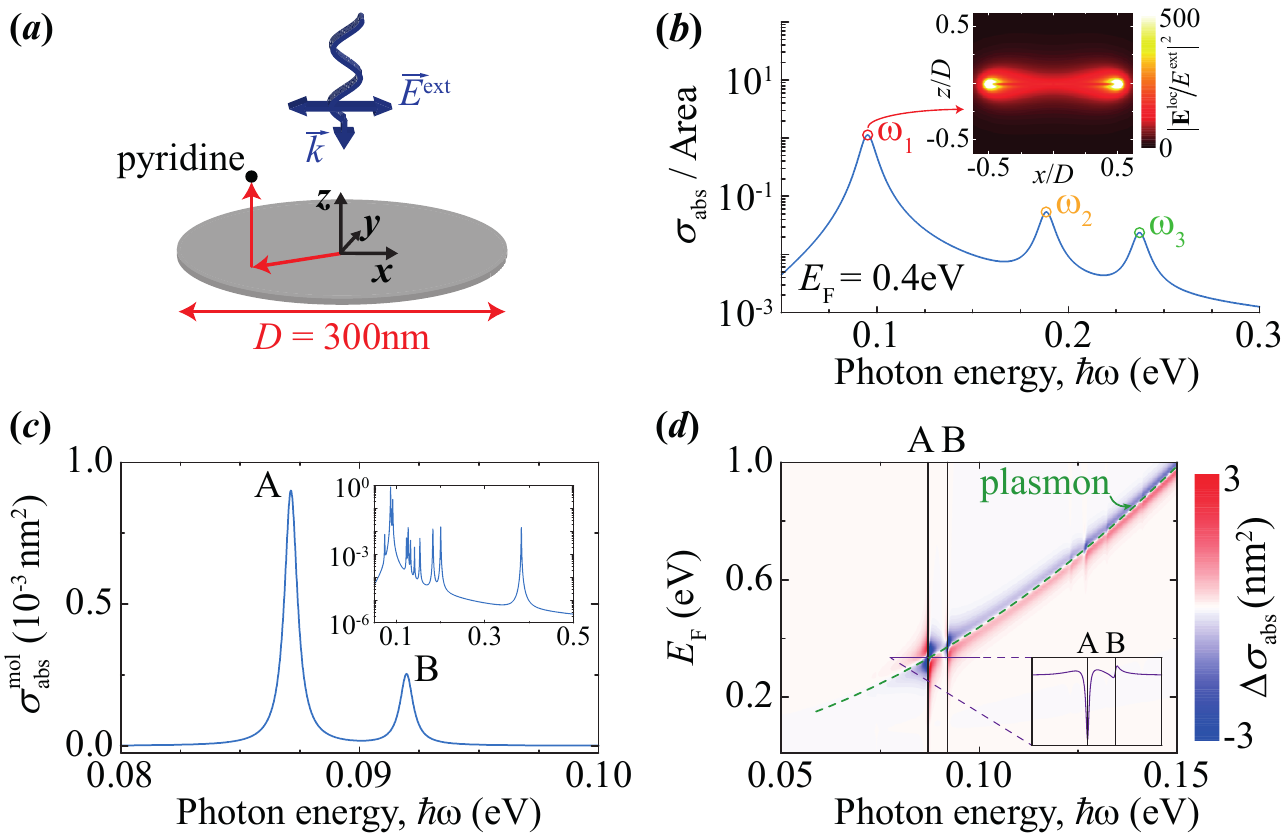}
\caption{{\bf Surface-enhanced infrared absorption (SEIRA) spectroscopy with graphene plasmons.} (a) Sketch of the structure considered in this work, consisting of a self-standing doped graphene nanodisk (diameter $D=300$\,nm) surrounded by target molecules. (b) Absorption cross-section spectrum of the nanodisk normalized to the graphene area for a Fermi energy $\EF=0.4$\,eV and mobility $\mu=2000$\,cm$^2/$\,Vs. The inset shows the near-field intensity of the lowest-energy dipolar plasmon in the $x-z$ plane. (c) Absorption cross-section of a pyridine molecule from available experimental data \cite{UWdatabase} (see Methods). (d) Change in the absorption cross-section induced by a single pyridine molecule placed near the disk edge at $(x,y,z)=(150\,$nm$,0,1\,$nm$)$ (see axes in Fig.\ \ref{Fig1}a), as a function of photon and graphene Fermi energies. The inset shows a line scan along a segment in the region where the main molecular absorption features A and B cross the lowest-order disk plasmon.}
\label{Fig1}
\end{center}
\end{figure*}

Infrared molecular vibrations lie in the spectral region where doped graphene structures with characteristic size in the range of tens to hundreds of nanometers support intense plasmons for attainable Fermi energies below 1\,eV. For simplicity, in what follows we focus on a self-standing graphene disk of diameter $D=300$\,nm (see Fig.\ \ref{Fig1}a). Following previously reported procedures \cite{paper176}, we simulate the response of the disk by solving Maxwell's equations using the boundary-element method \cite{paper040} (BEM), in which graphene is described as a disk of small thickness $t$ (we find convergence for $t=0.1\,$nm) and dielectric function $1+4\pi\ii\sigma(\omega)/\omega t$, where $\sigma(\omega)$ is the frequency-dependent surface conductivity of the carbon film. We use the local-RPA approximation for $\sigma(\omega)$ at room temperature \cite{paper235}, which also depends on the Fermi energy $\EF$ and the relaxation time $\tau$. The latter is estimated from the DC mobility as $\tau=\mu\EF/ev_{\rm F}^2$, where we assume a conservative value of $\mu=2000$\,cm$^2/$\,Vs. Notice that much larger mobilities have been measured in high-quality graphene \cite{NGM04}, and this parameter ultimately determines the energy resolution $\hbar\tau^{-1}$ that characterizes spectrometer-free sensing to determine molecular resonances (i.e., the sensing resolution is determined by the plasmon quality factor, and this is in turn proportional to the graphene mobility, as shown in Fig.\ \ref{Fig3}b below). The disk plasmons are clearly resolvable in the absorption spectrum (Fig.\ \ref{Fig1}b), exhibiting a dominant, low-energy feature of dipolar character (see upper inset to Fig.\ \ref{Fig1}b). It is important to realize that the plasmon frequencies scale as $\sim \sqrt{\EF}$, so they can be controlled by varying the Fermi energy through the addition of charge carriers to the system.

\begin{figure}
\begin{center}
\includegraphics[width=120mm,angle=0,clip]{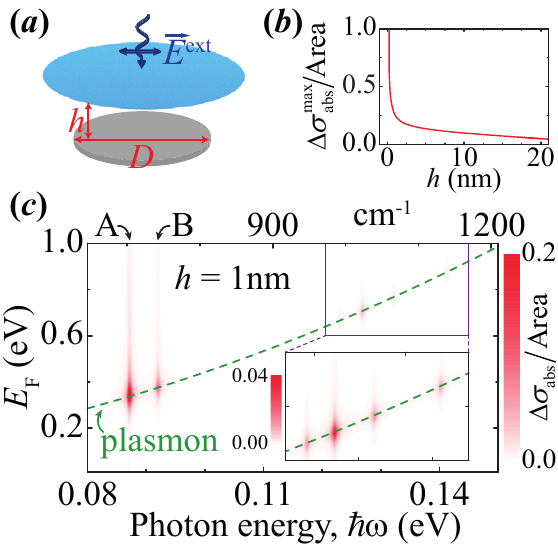}
\caption{{\bf Doping dependence of the absorption cross-section.} We represent in (c) the variation in the absorption cross-section produced by the interaction of a layer of pyridine molecules with the graphene disk considered in Fig.\ \ref{Fig1} (diameter $D=300\,$nm). The cross section is normalized to the disk area. The molecules are placed at a distance $h=1\,$nm above the carbon plane (see (a)) with a density of one molecule per nm$^2$ and covering an area that extends well beyond the disk edge. The lower-right inset to (c) shows a zoom of a high-photon-energy region in which weaker pyridine resonances are observable. Panel (b) shows the $h$ dependence of the absorption cross-section for photon and Fermi energies corresponding to the absolute maximum of the density plot of (b).}\label{Fig2}
\end{center}
\end{figure}

\subsection{SEIRA}

Infrared absorption is a first-order process in which the impinging infrared radiation excites roto-vibrational transitions allowed by the quantum selection rules.  We consider pyridine (${\rm C}_5{\rm H}_5{\rm N}$) as a generic molecule to illustrate this concept. The infrared absorption spectrum of pyridine is dominated by two experimentally observed lines \cite{UWdatabase} at photon energies $\hbar\omega_{\rm{A}}=0.087$\,eV and $\hbar\omega_{\rm{B}}=0.092$\,eV (see Fig.\ \ref{Fig1}c and Methods). Upon external excitation by a normal-incidence light plane wave of frequency $\omega$ and electric field amplitude $\Eb^{\rm ext}$, the molecule experiences an enhanced local field $\Eb^{\rm loc}=\Eb^{\rm ext}+\Eb^{\rm ind}+\Eb^{\rm self}$, given by the superposition of $\Eb^{\rm ext}$, the field induced by the nanodisk $\Eb^{\rm ind}$, and the self-induced field of the molecule $\Eb^{\rm self}$. The latter is negligibly small for most molecules, so we disregard it in what follows. The enhancement in the SEIRA cross-section is simply described by the increase in local field intensity, which we approximate as $|\Eb^{\rm ind}/E^{\rm ext}|^2$. This quantity, which reaches large values, as shown in the inset to Fig.\ \ref{Fig1}b, coincides with $|G|^2$, where $G$ is given by the semi-analytical expression of Eq.\ (\ref{G}) (see Methods). The resulting enhancement is represented in Fig.\ \ref{Fig1}d, where we plot the change in the absorption cross-section of a graphene disk by placing a single pyridine molecule near its edge. The variation in the cross section $\Delta\sigma_{\rm abs}$ is clearly enhanced along the dipolar plasmon line, which shifts in energy with varying doping level as noted above. Importantly, $\Delta\sigma_{\rm abs}$ reaches values that are three orders of magnitude larger than the cross section of the isolated molecule.

\begin{figure}
\begin{center}
\includegraphics[width=160mm,angle=0,clip]{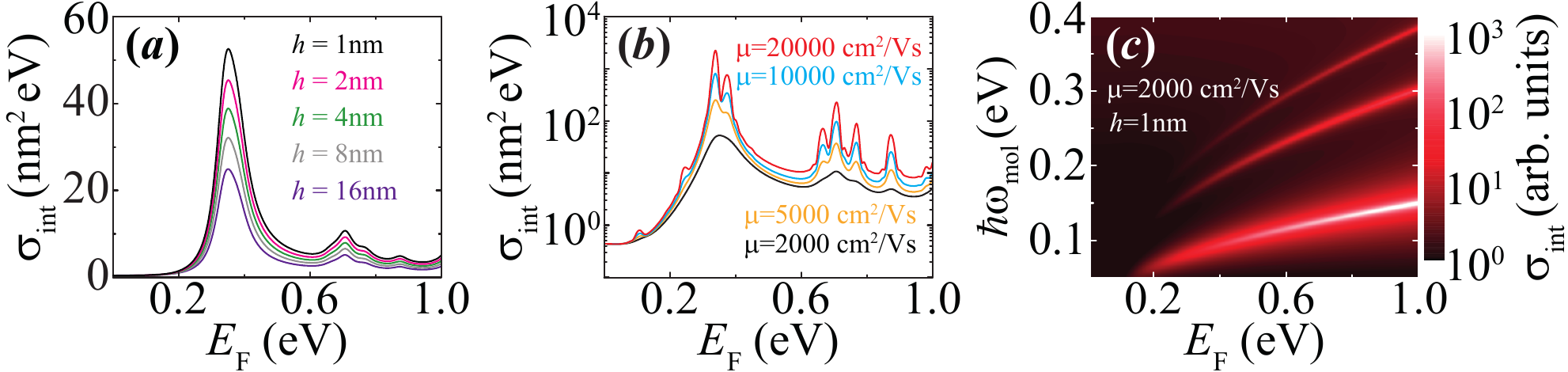}
\caption{{\bf Molecular sensitivity of the doping-dependent frequency-integrated absorption.} (a,b) We analyze the integral over photon energies ($0-1\,$eV) of the change in absorption cross-section (i.e., $\sigma_{\rm int} = \hbar \int d\omega\,\Delta\sigma_{\rm abs}$) produced by a layer of pyridine molecules as a function of graphene Fermi energy $\EF$ for (a) several molecule-graphene distances with fixed graphene mobility $\mu=2000$\,cm$^2/$\,Vs, and (b) fixed distance $h = 1$\,nm and different values of $\mu$. (c) We analyze $\sigma_{\rm int}$ for $\mu=2000$\,cm$^2/$\,Vs and $h = 1$\,nm with a generic molecule exhibiting a single absorption resonance at frequency $\omega_{\rm mol}$ with fixed maximum polarizability and bandwidth $\hbar\Delta\omega_{\text{mol}}=1$\,meV. The integrated absorption is depicted as a function of $\EF$ and $\hbar\omega_{\rm mol}$.}
\label{Fig3}
\end{center}
\end{figure}

In practical applications, one is interested in sensing a small concentration of molecules placed over a range of distances from the graphene disk. For reference, we consider a monolayer of pyridine molecules separated 1\,nm from the graphene and with a density of one molecule per nm$^2$. As shown in Fig.\ \ref{Fig2}b and Fig.\ \ref{Fig3}a, the results are qualitatively similar over distances up to a few nanometers from the graphene, and therefore, the technique should be robust against the uncertainty in the exact location of the molecules, provided their separation is in the $<10\,$nm range. The change in the absorption cross-section due to the molecular layer is simply obtained by integrating the change due to a single molecule $\Delta\sigma_{\rm abs}$ over the layer area and multiplying by the molecule surface density. The result is shown in Fig.\ \ref{Fig2}c as a function of photon and graphene Fermi energies, $\hbar\omega$ and $\EF$, respectively. The important message from this plot is that there is a one-to-one correlation between the molecular resonance photon energies and the values of $\EF$ at which the nanodisk dipolar plasmon band overlaps with those resonances. More precisely, this leads to peaks in the absorption cross-section at $\EF = 0.34$\,eV and $0.39$\,eV, corresponding to molecular resonances of energies $\hbar\omega_{\rm A}$ and $\hbar\omega_{\rm B}$, respectively.

The noted one-to-one correspondence between molecular resonances and Fermi energies suggests that it is possible to obtain spectral information by recording the absorption as a function of $\EF$, rather than $\hbar\omega$. Indeed, upon illumination by spectrally broad sources (e.g., an infrared lamp), this graphene-based sensor device can discriminate resonant photon energies by examining the Fermi levels at which the measured (spectrally unresolved) absorbed power is peaked. The disk plasmons act as amplifiers of the incident light at gate-controlled photon energies. We illustrate this concept by calculating the integral over photon energies ($0-1\,$eV) of the change in absorption cross-section, $\sigma_{\rm int} = \hbar \int d\omega\,\Delta\sigma_{\rm abs} $. Results for several molecule-layer/graphene distances  (Fig.\ \ref{Fig3}a) suggest that the sensor can perform similarly well up to a distance $\sim10\,$nm. Additionally, we explore a range of  feasible graphene mobilities (Fig.\ \ref{Fig3}b), ranging from the conservative value that we use in Fig.\ \ref{Fig1}, to higher-quality graphene. Although the former is already capable of giving sufficient molecule-specific information to resolve the presence of pyridine, we note that currently attainable high-quality graphene enables further discrimination of weak vibrational features (e.g., it allows us to resolve the A and B resonances, which are separated by $\sim5$\,meV). We further illustrate the underlying principle of the proposed sensing scheme by considering a generic molecule that exhibits an absorption resonance at frequency $\omega_{\rm mol}$, with fixed bandwidth $\hbar\Delta\omega_{\rm mol}=1$\,meV and fixed maximum polarizability.
The molecule is uniformly distributed on a plane at a distance of 1\,nm from the graphene surface and we considering a moderate mobility of $\mu=2000$\,cm$^2/$\,Vs. We integrate the absorption cross-section over the $\hbar\omega<1\,$eV spectral range. The resulting integrated cross-section (Fig.\ \ref{Fig3}c) displays clear maxima when the graphene plasmons are on resonance with the molecule. For a given molecular resonance, we have to consider a horizontal cut over the density plot, from which $\omega_{\rm mol}$ can be retrieved by looking at $\EF$ for maximum absorption. Incidentally, the combination of the first three dipole-active disk plasmon resonances allows us to cover a broad molecular spectral range. As shown in Fig.\ \ref{Fig3}b, the {\it spectral} resolution of this spectrometer-free technique is limited by the plasmon width $\sim\hbar\tau^{-1}$, which is in turn dominated by impurity scattering. Incidentally, we observe a moderate increase in the integrated cross-section by changing the mobility.

\begin{figure*}
\begin{center}
\includegraphics[width=160mm,angle=0,clip]{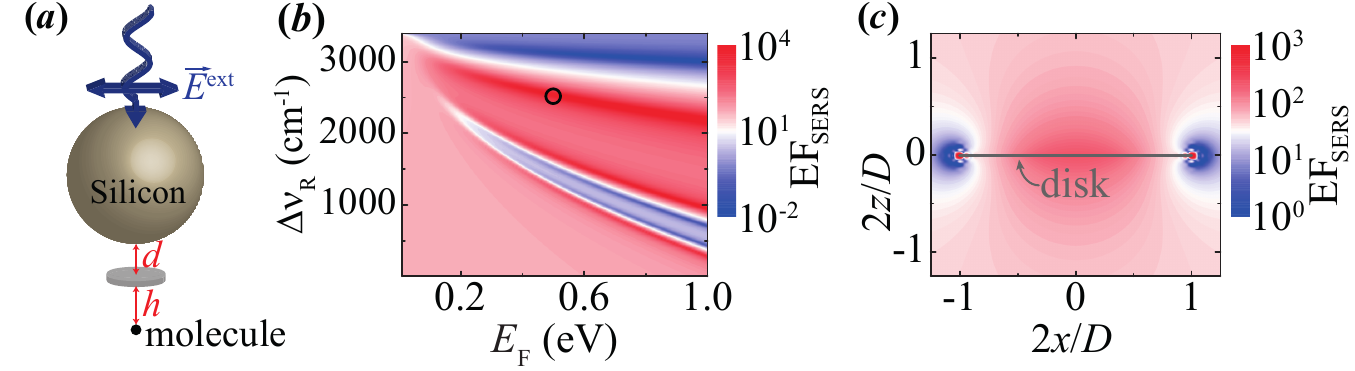}
\caption{{\bf Surface-enhanced Raman scattering (SERS) with graphene plasmons.} (a) Sketch of the system under consideration, consisting of a silicon sphere (diameter $1530$\,nm, $\epsilon_{\rm Si}=12$) placed a distance $d=9\,$nm above a graphene disk (diameter $D=300$\,nm), which is in turn placed at a distance $h=1\,$nm above a Raman active molecule. The system is irradiated with a $0.422$\,eV (i.e., $2.94$\,$\mu\rm{m}$ wavelength) light plane wave that is resonant with a Mie mode of the sphere (i.e., the sphere works as a nanofocuser, similarly to previous designs \cite{MGK12}). (b) SERS enhancement factor ${\text{EF}}_{\text{SERS}}$ (relative to an isolated molecule), as a function of Raman-shift $\Delta\nu_{\text{R}}$ and graphene Fermi energy $\EF$ for a molecule placed along the axis of symmetry. (c) SERS enhancement as a function of the position of the molecule relative to the graphene disk for doping and Raman shift conditions corresponding to the open circle in (b).}
\label{Fig4}
\end{center}
\end{figure*}

\subsection{SERS}

Raman scattering is a second-order process that is proportional both to the incident light intensity and to the emission from the inelastically frequency-shifted transition dipole. Unfortunately, owing to the non-resonant nature of this effect, the cross sections of single Raman-active molecules are very small ($\sim10^{-26}$\,cm$^2$). Following a similar strategy as above, we use graphene plasmons to amplify the emission, and we supplement the structure with a resonant silicon cavity \cite{MGK12} that also amplifies the incident light intensity (Fig.\ \ref{Fig4}a). The resulting SERS enhancement with respect to the isolated molecule becomes
\begin{equation}
{\text{EF}}_{\rm SERS} = \left|\frac{\Eb^{\rm loc}}{E^{\rm ext}}\right|^2 \left| \frac{\pb^{\rm tot}}{p^0} \right|^2 ,
\nonumber
\end{equation}
where $\Eb^{\rm loc}$ is local field at the incident light frequency, mainly controlled by the silicon cavity, whereas $\pb^{\rm tot}=\pb^0+\pb^{\rm ind}$ is the total superposition of the free-molecule Raman transition dipole moment $\pb^0$ and the dipole induced by the molecule on the surrounding structure. For simplicity, we calculate $\Eb^{\rm loc}$ considering only the silicon cavity and neglecting the graphene, while we obtain $\pb^{\rm ind}$ as the dipole induced on the graphene disk without taking into account the silicon cavity. Like in SEIRA, we neglect the self-induced polarization of the molecule, and in this approximation, the disk response enters through the same coupling coefficient $G$ in both SEIRA and SERS (see Eq.\ (\ref{G}) in the Methods section).

The calculated enhancement factor $\rm EF_{SERS}$ is plotted in Fig.\ \ref{Fig4}b as a function of the molecule Raman-shift $\Delta\nu_{\rm R}$ and the graphene Fermi level $\EF$ for a pump photon energy $\hbar\omega_{\rm pump}=0.422$\,eV (i.e., the emission line of an Er:YAG laser), on resonance with the silicon cavity, which produces an enhancement $\left|\Eb^{\rm loc}/E^{\rm ext}\right|^2\approx2200$. The resulting $\rm EF_{SERS}$ reaches $\sim10^4$ (see Fig.\ \ref{Fig4}b), and its actual value depends on the position of the molecule relative to the disk (see Fig.\ \ref{Fig4}c), yielding again qualitatively similar performance for molecule-graphene distances in the $<10\,$nm region.

In order to evaluate the possibility of realizing a spectrometer-free SERS sensor, we consider the integral of the SERS intensity over Raman shifts in the $0-1\,$eV range. Additionally, as the molecule position in a realistic experiment cannot be controlled, we also average over molecules distributed on a layer placed 1\,nm below the graphene and extending beyond the disk area. Using the relation between SERS and SEIRA through the same disk-coupling coefficient $G$ (see Methods), we find exactly the same result as in Fig.\ \ref{Fig3}c, up to a uniform multiplicative factor, for the distribution of the SERS intensity as a function of the graphene Fermi energy $\EF$ and the Raman shift $\omega_{\rm pump}-\omega_{\rm mol}$ (i.e., just replace the vertical axis label in Fig.\ \ref{Fig3}c by the Raman shift). Obviously, for fixed $\omega_{\rm pump}$ the Raman shift depends on the molecular vibration frequency $\omega_{\rm mol}$. We note again the presence of three prominent emission features arising from the three dominant plasmon modes excited in the graphene nanodisk within the spectral range under consideration. For each plasmon mode, there is a one-to-one correspondence between $\EF$ and $\hbar\omega_{\rm mol}$ in the photon-energy-integrated SERS intensity. Overall, upon electrical tuning of the graphene, the three modes span an energy range of $\sim0.4$\,eV, which is larger than the individual plasmon resonance widths, and thus, broadband SERS is achievable.

\section{CONCLUDING REMARKS}

An important practical aspect of the proposed sensor is the control of the graphene Fermi energy. We envision a gating device in which a bottom gate is combined with a contact for the graphene. Electrical connectivity could be provided through a thin transparent insulating layer, as recently used to demonstrate active control of graphene disk plasmons \cite{paper212}. Alternatively, we expect similar results for graphene ribbons, whose plasmon frequencies and characteristics for transversal polarization are similar to those of the disks, with the advantage that these structures can be contacted in a region far from the active sensing area.

The change in Fermi energy produced by the target molecules can be a serious problem that might limit the applicability of the proposed sensing technique, as it adds an element of uncertainty in the determination of the graphene neutrality point. We anticipate several possible strategies to deal with this uncertainty: (1) The entire spectrum changes when moving $\EF$, and therefore, it should be sufficient to resolve spectral distances associated with the molecular features (i.e., we can adapt the above analyses to resort on the dependence on $\EF^2$, rather than on $\EF$, as the former is linear with the applied voltage, and therefore, the problem reduces to determining the offset of this voltage for each analysis). (2) In many practical situations, one is interested in discriminating between a certain finite number of different detected molecules, so the proposed sensor can be calibrated for each of them, including the noted charge-transfer effect. (3) Finally, charge transfer can be drastically reduced through the addition of a thin transparent insulating layer, which according to Fig.\ \ref{Fig3}a can have a thickness of several nanometers without causing a serious reduction in sensing capabilities.

In conclusion, our calculations clearly show the ability of graphene to resolve the chemical identity of adsorbed molecules from the measurement of broadband-integrated absorption and Raman scattering signals enhanced by the electrically tunable plasmons of this material. The narrowness of these plasmons is sufficient to resolve the frequency of the molecular resonances in the integrated intensity as a function of doping Fermi energy. We thus propose a new infrared sensing strategy that avoids the use of costly and inefficient optical elements in this frequency regime, such as spectrometers and laser sources, and instead simply involves infrared lamps and electrical doping of the graphene structure through an externally applied gate voltage. The large confinement and electric-field amplification associated with graphene plasmons leads to SEIRA and SERS enhancements reaching $\approx10^3$ and $\approx10^4$, respectively, which also suggest the use of graphene to improve traditional sensing techniques based on spectrally resolved infrared absorption and Raman scattering. The vibrational-energy resolution of the proposed sensing scheme is determined by the spectral width of graphene plasmons (i.e., it is essentially limited by material quality) and can reach a few meV under currently attainable conditions. In summary, graphene plasmons provide a versatile platform for sensing, opening new possibilities by exploiting the large electro-optical tunability of this material, and in particular, the realization of label-free chemical identification without the involvement of laser sources and spectrometers.


\begin{table}
\centering
\begin{tabular}{  c  c  c  }\hline\hline \\
  $m$  & $\hbar\omega_m(\rm eV)$ & $\alpha_m\left(10^{-3} \r{A}^3\,\,{\rm eV}^2\right)$  \\ & & \\ \hline & & \\
  $1$  &     $0.074$    &          $0.073$               \\
  $2$  &     $0.087$    &          $9.328$               \\
  $3$  &     $0.092$    &          $2.591$               \\
  $4$  &     $0.123$    &          $0.052$               \\
  $5$  &     $0.127$    &          $0.155$               \\
  $6$  &     $0.132$    &          $0.042$               \\
  $7$  &     $0.141$    &          $0.021$               \\
  $8$  &     $0.153$    &          $0.052$               \\
  $9$  &     $0.183$    &          $0.135$               \\
  $10$ &     $0.200$    &          $0.166$              \\
  $11$ &     $0.382$    &          $0.156$              \\ & & \\ \hline\hline
\end{tabular}
\caption{Fitting parameters for the polarizability of pyridine (see Eq. (\ref{alpha})).}
\label{Table1}
\end{table}

\section{METHODS}

\begin{figure}
\begin{center}
\includegraphics[width=110mm,angle=0,clip]{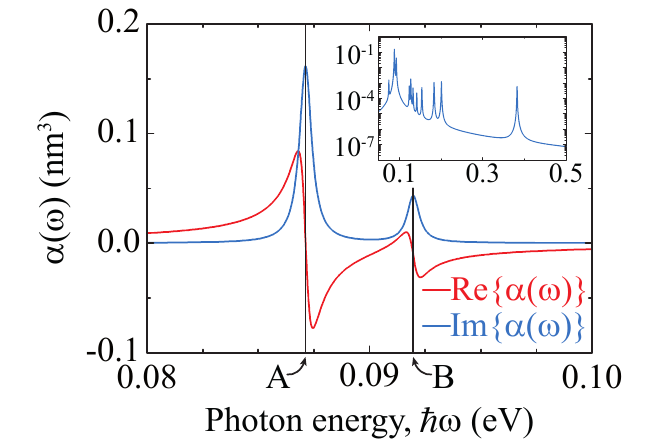}
\caption{Real (red curve) and imaginary (blue curve) parts of the polarizability of pyridine in the spectral range here considered, dominated by the absorption lines A and B. The inset shows the imaginary part over a wider energy region.}
\label{Fig5}
\end{center}
\end{figure} 

\subsection{Optical response of Pyridine}

We extract the polarizability of pyridine by fitting available experimental data \cite{UWdatabase} to a sum of Lorentzians (automatically satisfying Kramers-Kronig causality relations),
\begin{equation}
\alpha(\omega) = \sum_m \frac{\alpha_m}{\omega_m^2-\omega(\omega+\ii\gamma)},
\label{alpha}
\end{equation}
where $\alpha_m$ and $\omega_m$ are fitting parameters (see Table\ \ref{Table1}) and we assume a fixed bandwidth $\hbar\gamma = 0.7$\,meV. Two of these Lorentzians ($m=2,3$) dominate the spectral range here considered (see Fig.\ \ref{Fig5}), whereas additional terms are needed to describe a broader range of energies (inset to Fig.\ \ref{Fig5}). We use Eq. (\ref{alpha}) in the calculations of Figs.\ \ref{Fig1}, \ref{Fig2}, and \ref{Fig3}a,b.

\begin{figure}
\begin{center}
\includegraphics[width=160mm,angle=0,clip]{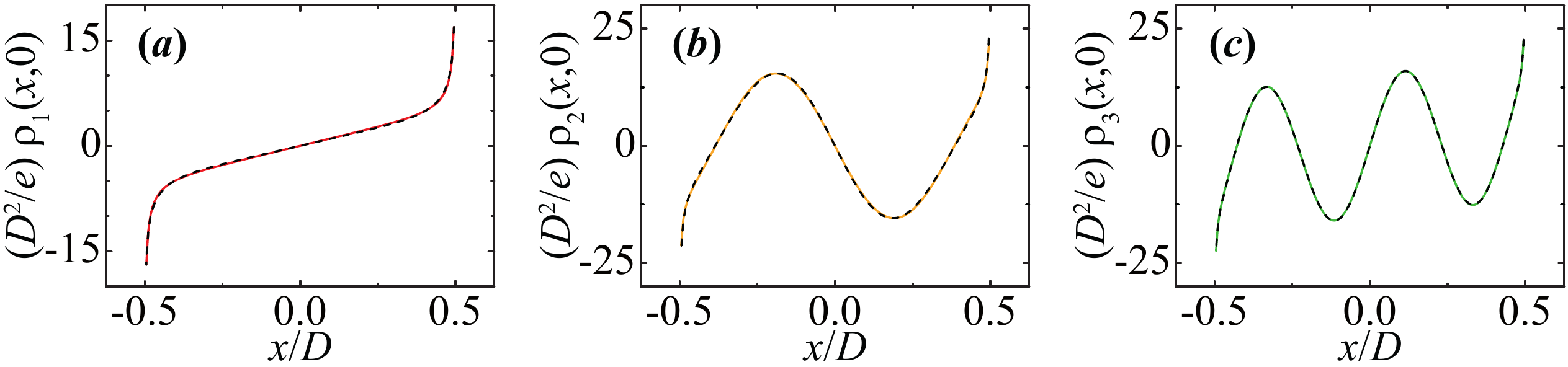}
\caption{Charge densities $\rho_j$ associated with the dominant dipolar plasmons of a graphene disk $j=1-3$ as a function of the coordinate $x$, parallel to the direction of polarization. The axes are normalized using the disk diameter $D$. Full numerical simulations (solid curves) are compared with the analytical expressions given in the text (broken curves).}
\label{Fig6}
\end{center}
\end{figure}

\subsection{Optical response of the graphene disk}

The response of the disk is simulated by numerically solving Maxwell's equations, as discussed above. Additionally, in order to reduce the computational cost of exploring the wide range of parameters discussed in this work, we use a simple semi-analytical method that yields indistinguishable results on the scale of the figures. More precisely, as the disk diameter $D$ is much smaller than the light wavelength, we work in the electrostatic limit and adopt an eigenmode expansion \cite{paper228,paper235} that allows us to express the response in terms of a few dominant plasmons. A simple extension of this method leads to an expansion of the induced density as a sum over different plasmon modes,
\[\rho^{\rm ind}(\Rb)=\frac{D}{e}\sum_j\frac{c_j}{\ii\omega D/\sigma+1/\eta_j}\rho_j(\Rb),\]
where $\rho_j$ is the induced charge associated with plasmon $j$ of eigenvalue $\eta_j$, the coordinate vector $\Rb$ runs over the graphene area, $\sigma$ is the surface conductivity, and  the expansion coefficients are related to the externally applied potential through
\[c_j=\frac{1}{e}\int d^2\Rb\;\rho_j(\Rb) \phi^{\rm ext}(\Rb).\]
For SEIRA, we write the external potential as $\phi^{\rm ext}(\Rb)=-\Rb\cdot\Eb^{\rm ext}$ and evaluate the field induced by the disk at the molecule position $\rb_0$ as $\Eb^{\rm ind}=\int d^2\Rb\;\rho^{\rm ind}(\Rb)\;(\rb_0-\Rb)/|\rb_0-\Rb|^3$. For SERS, the external potential produced by the Raman emission dipole $\pb^0$ is $\phi^{\rm ext}(\Rb)=-\pb^0\cdot(\rb_0-\Rb)/|\rb_0-\Rb|^3$, whereas the dipole induced on the disk reduces to $\pb^{\rm ind}=\int d^2\Rb\;\rho^{\rm ind}(\Rb)\;\Rb$. For simplicity, we take the light field and the molecule dipole both oriented along $x$, so we can write the relevant components of the fields as $\Eb^{\rm ind}=G\,\Eb^{\rm ext}$ and $\pb^{\rm ind}=G\,\pb^0$ in terms of the dimensionless coefficient
\begin{equation}
G=\frac{-D}{e^2}\sum_j\frac{1}{\ii\omega D/\sigma+1/\eta_j}\left(\int d^2\Rb\,x\rho_j(\Rb)\right)\left(\int d^2\Rb'\,\frac{(x_0-x')\rho_j(\Rb')}{|\rb_0-\Rb'|^3}\right).
\label{G}
\end{equation}
In practice, only three plasmons are necessary to describe the frequency range under consideration, with eigenvalues $\eta_1=-0.0664$, $\eta_2=-0.0162$, and $\eta_3=-0.0099$, as obtained from the frequencies $\omega_j$ shown in Fig.\ \ref{Fig1}b. We calculate the corresponding charge densities using BEM and we obtain the results shown in Fig.\ \ref{Fig6}, which can be approximated through the fitting analytical expressions
\begin{align}
\rho_1(\Rb)&=10.13\,\theta\;g(\Rb), \nonumber\\
\rho_2(\Rb)&=12.11\,[\theta+1.08\,\sin(9.1\,\theta)]\;g(\Rb), \nonumber\\
\rho_3(\Rb)&=37.15\,[\theta-22.5\,\theta^3+75.8\,\theta^5-0.5\,\sin(13.5\,\theta)]\;g(\Rb),\nonumber
\end{align}
where $\theta=R/D$ and \[g(\Rb)=(e/D^2)\;\cos\varphi\;\left[1+\exp[-5(1-2\theta)]/\left(4\sqrt{1-2\theta}\right)\right]\] and we use polar coordinates $\Rb=(R,\varphi)$.

\section*{Acknowledgement}

This work has been supported in part by the European Commission (Graphene Flagship CNECT-ICT-604391 and FP7-ICT-2013-613024-GRASP).


\end{document}